\documentclass[twocolumn,aps,amsmath,amssymb,showpacs,pre]{revtex4}
\usepackage{graphicx}
\usepackage{epsfig}
\bibstyle{apsrev.bib}

\begin{document}
\input epsf.sty

\title{Superdiffusion in a class of networks with marginal long-range connections}

\author{R\'obert Juh\'asz}
 \email{juhasz@szfki.hu} 
\affiliation{Research Institute for Solid
State Physics and Optics, H-1525 Budapest, P.O.Box 49, Hungary}

\date{\today}

\begin{abstract}
A class of cubic networks composed of a 
regular one-dimensional lattice and a set of long-range links is introduced. 
Networks parametrized by a positive integer $k$ are constructed 
by starting from a one-dimensional lattice and iteratively connecting
each site of degree $2$ with a $k$th neighboring site of degree $2$.
Specifying the way pairs of sites to be connected are selected, various
random and regular networks are defined, all of which 
have a power-law edge-length distribution of the 
form $P_>(l)\sim l^{-s}$ with the marginal exponent $s=1$.  
In all these networks, lengths of shortest paths 
grow as a power of the distance and random walk is
super-diffusive.   
Applying a renormalization group method, the corresponding 
shortest-path dimensions 
and random-walk dimensions are calculated exactly
for $k=1$ networks and for $k=2$ regular networks; 
in other cases, they are estimated by numerical methods.
Although, $s=1$ holds for all representatives of this class, 
the above quantities are found to depend on the details of the
structure of networks controlled by $k$ and other parameters. 
  
\end{abstract}


\maketitle

\newcommand{\bc}{\begin{center}}
\newcommand{\ec}{\end{center}}
\newcommand{\be}{\begin{equation}}
\newcommand{\ee}{\end{equation}}
\newcommand{\beqn}{\begin{eqnarray}}
\newcommand{\eeqn}{\end{eqnarray}}

\vskip 2cm
\section{Introduction}

Random walk in disordered environments is a much studied problem as it is
a basic model in the theory of transport in heterogeneous
media \cite{havlin,bouchaud} and it has features 
much different from those characteristic of homogeneous systems. 
In many cases, the diffusion law is altered in the way that the 
typical displacement of the random walker still grows 
as a power of time, i.e. $R(t)\sim t^{\nu}$,
however, the diffusion exponent $\nu$ differs from 
the value $\nu=1/2$ characteristic of normal diffusion.  
In case of random walks with quenched random jump rates on regular
lattices or random walks on
fractal lattices (e.g. percolation clusters) the process is 
in general sub-diffusive, i.e. $\nu<1/2$ \cite{havlin,bouchaud}. 
The opposite case, $\nu>1/2$, where the particle is speeded
up compared to normal diffusion, is much rarer.  
This phenomenon, called superdiffusion, 
arises e.g. in turbulent fluids \cite{swk}, 
in chaotic transport in laminar fluid flows \cite{solomon} or in
systems of polymerlike breakable micelles \cite{oblu}.   
In random walk models, superdiffusion can be induced by allowing long-range
jumps: In the L\'evy flight \cite{bouchaud}, jumps of arbitrary length 
are executed with a probability that decays as a power of the length; 
in its ``discretized'' version, the Weierstrass walk, 
the possible jump lengths are integer powers of an integer $a>1$ 
\cite{weierstrass}.      
     
Another possibility for superdiffusion in case of random walks on a lattice is 
when long-range jumps occur only at certain sites of the
lattice. In most cases, the underlying lattice is a union of a
regular and a random graph, where the latter has arbitrarily long edges. 
Such graphs have been investigated in various aspects. 
As a variant of the small-world network model constructed by
rewiring edges of a regular network \cite{ws}, 
Newman and Watts have considered a one-dimensional lattice to 
which ``shortcuts'' between randomly chosen pairs of sites are 
added \cite{nw}. 
In this model, shortest paths \cite{nw,moukarzel}, as well as   
spectral properties of the Laplacian \cite{monasson} have been studied. 
Shortest paths or the diameter of d-dimensional lattices where 
long edges exist between pairs of sites with probabilities that 
decay algebraically with the distance have been studied
by several authors 
for $d=1$ \cite{jespersen,bb,sc,mam} and for $d\ge 1$ \cite{coppersmith}.
Beside the above models, $n$-regular graphs of this type,
where the degree of all nodes (i.e. the number of edges emanating from a
node) is $n$, have been introduced, as well. 
The issue of decentralized algorithms for finding short paths has been
considered by Kleinberg in a $d$-dimensional lattice where each
site has a fixed number of directed edges to randomly chosen sites
that are selected with power-law decaying probabilities \cite{kleinberg}.      
Benjamini and Hoffman have studied minimal paths in $\omega$-periodic
graphs, which are unions of periodic graphs over the integers \cite{bh}.
Recently, Boettcher et al. have introduced a hierarchical $3$-regular 
network consisting of a
one-dimensional lattice and recursively constructed long-range links 
and calculated the random-walk dimension $d_{\rm rw}\equiv 1/\nu$ 
of this network \cite{boettcher}.   

In this paper, we shall consider networks composed of a one-dimensional 
lattice and an additional set of long links. 
Particularly, we focus on the intriguing situation when the tail of
the distribution of edge-lengths is of the form $P_>(l)\sim \beta l^{-1}$.
In this case, the diameter is conjectured to grow algebraically with
the size of the network with a non-universal $\beta$-dependent
exponent \cite{bb,coppersmith}. 
The aim of this work is to confirm this conjecture by explicit
calculations and to probe whether the diffusion exponent in such
networks displays a similar ``marginal'' behavior.
For this purpose, we define a class of cubic
(i.e. $3$-regular) networks with marginal edge-length distributions 
(i.e. $P_>(l)\sim l^{-1}$)  and shall 
constructively demonstrate that the shortest-path dimension
that characterizes the size-dependence of the diameter and the 
diffusion exponent are not exclusively determined by the power 
in the edge-length distribution but 
depend on the details of the structure of networks.
The above two intrinsic properties of networks 
are calculated exactly in certain cases
by means of a renormalization group method; in other cases, they are
estimated by the numerical implementation of the renormalization 
procedure and by numerical simulation. 

The rest of the paper is organized as follows.
In Section \ref{model}, 
a class of networks parametrized by a positive integer $k$ is defined
and its general features are discussed. In Section \ref{sectionk1},
networks with $k=1$ whereas in Section \ref{sectionk2}, networks with $k>2$
are investigated in detail. Results are discussed in Section \ref{disc} and a
heuristic derivation of the relation to resistor networks is 
presented in the Appendix.   

\section{General features}
\label{model}
  
\subsection{Construction of the networks}

The networks to be studied in the subsequent sections have in common 
that they are all constructed in the following way. 
A one-dimensional open or periodic lattice with $N$ sites is given, where
sites are numbered consecutively from $1$ to $N$.   
The degree of all sites is thus initially $2$ in the periodic lattice
and all but site $1$ and $N$ in the open one.
Sites of degree $2$ will be called in brief {\it active sites}. 
The links of this initial regular lattice will be termed 
{\it short links} in the followings in order to distinguish them from  
{\it long links} generated by the following procedure.    
Let us assume that $N$ is even and $k$ is a fixed
positive integer.   
A pair of active sites is selected such that the number of 
active sites between them is $k-1$ and this pair is 
then connected by a (long) link. 
That means, for $k=1$, neighboring active sites are connected, for $k=2$
next-to-neighboring ones, etc.
This step, which renders two active sites to sites of degree $3$, 
is then iterated until $2(k-1)$ active sites are left. 
These are then paired in an arbitrary way,    
which does not affect the exponents appearing in asymptotic relations in 
the limit $N\to\infty$. 
In the resulting network, all sites are of degree $3$ if the procedure
starts from a periodic lattice whereas, in case of an open
lattice, site $1$ and site $N$ remain of degree $1$.  
The networks generated in this way (or, more precisely, 
the ensembles of networks in case of random networks) 
are characterized by the number $k$ and by the way pairs are selected. 

At some stage of the construction procedure,
when the number of active sites is $N_a$,
these sites are distributed 
homogeneously on a coarse-grained scale $\xi\gg 1/c$, 
where $c\equiv N_a/N$ is their number density.
In other words, 
spacings between neighboring active sites have a rapidly decaying
distribution with the expected value $1/c$. 
Thus, long links of length larger than $l$ are generated typically when $c$ is
smaller than $1/l$ and we obtain for the distribution of
lengths in an infinite network ($N\to\infty$): 
$P_>(l)\sim l^{-1}$. 
Disregarding regular networks (see Section \ref{aper} and
\ref{aperk2}), where exclusively long links of length 
$l_n\sim\lambda^n$  ($\lambda>1$, $n=1,2,\dots$) form, 
the probability of a long link of length $l$ is thus inversely
proportional to the square of the length: $p_l\sim l^{-2}$. 
Note, however, that not all edge lengths are realized in the
construction procedure even for random networks.
Namely, for $k=1$, only long links of odd length are produced.   

\subsection{Studied quantities}

We are interested in two intrinsic properties of networks.
Beside the distance $l$ measured on the underlying 
one-dimensional lattice, we also consider another metric:  
The chemical distance (or the
length of the shortest path) $\ell$ between two sites 
is the minimum number of links that have to be
traversed when going from one site to the other.
We will see that, in the networks under study, the average 
length of shortest path between sites located in a distance $l$ grows
algebraically with $l$ for large distances: 
$\overline{\ell(l)}\sim l^{d_{\rm min}}$, 
where the shortest-path dimension $d_{\rm min}<1$ is characteristic 
of the particular network. 
This dimension describes at the same time the finite-size scaling of
the diameter $D(N)$ of a network, 
which is the maximum of chemical distances
between any pairs of sites: $D(N)\sim N^{d_{\rm min}}$.
 
The other quantity of interest is the random-walk dimension of the
network \cite{gaa,havlin}. 
We consider a continuous time random walk on (infinite) 
networks, where the walker can jump 
with unit rate to any of the sites connected with the site it resides.
The random-walk dimension $d_{\rm rw}$ is defined through the relation 
$[\langle x^2(t)\rangle ]_{\rm typ}\sim t^{2/d_{\rm rw}}$,
where $x(t)$ denotes the displacement of the walker (measured 
on the underlying one-dimensional lattice) at time $t$ and  
$[\langle x^2(t)\rangle ]_{\rm typ}\equiv\exp{\overline{\ln\langle x^2(t)\rangle}}$ is the ``typical value'' of $\langle x^2(t)\rangle$.   
Here, $\langle\cdot\rangle$ denotes the expected value for a fixed starting position in a fixed realization of the ensemble of networks, while the overbar stands for the average
over starting positions and the ensemble of networks.  
Note that the expected value $\overline{\langle x^2(t)\rangle}$   
does not exist if $t>0$ since the expected value of edge lengths
is infinite (in infinite networks). This accounts for that the average of 
$\ln\langle x^2(t)\rangle$ is considered instead. 
The practical reason of the second averaging procedure is to eliminate 
random modulations which stem from the randomness
of the structure of networks, whereas for regular networks, this
averaging may be ignored.    

\subsection{Relation to resistor networks}

The calculation of random-walk dimension is based on
the well-known relation between the effective diffusion constant and
the effective resistance of the equivalent resistor network 
\cite{bouchaud,havlin}. 
In the equivalent resistor network, 
each link has a (dimensionless) resistance $r_i$
related to the jump rate $w_i$ along that link via $r_i=1/w_i$.  
Considering networks built on an open lattice, 
the time $t$ needed for the walker to get from
one end of the network (site $1$) to the other one (site $N$) is 
\be 
t\sim N\tilde r
\label{resistor}
\ee
for large $N$, where $\tilde r$ is the effective 
resistance of the equivalent resistor network between the two endpoints. 
For the sake of self-containedness, a heuristic derivation of
this relation is given in the Appendix; 
for a precise formulation of this connection in an
arbitrary network, the reader is referred to Ref. \cite{gefen}.    
Eq. (\ref{resistor}) implies that the random-walk dimension is related
to the resistance exponent $\zeta$ defined by the 
asymptotical relation $\tilde r(N)\sim N^{\zeta}$ as  
\be 
d_{\rm rw}=1+\zeta.
\label{dwzeta}
\ee
So, the problem of calculating $d_{\rm rw}$ is reduced to the calculation of
the resistance exponent $\zeta$ of the equivalent resistor network.

\section{Networks with $k=1$}
\label{sectionk1}
   
\begin{figure}[h]
\includegraphics[width=0.9\linewidth]{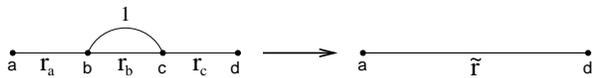}
\caption{\label{fig1} Replacement of a minimal loop by a single
  link.}
\end{figure}

The simplest class of networks is obtained when neighboring active 
sites are connected in the construction procedure, i.e. $k=1$. 
In this case, the elementary step of the calculation of the
resistance exponent is that a minimal loop 
(pair of sites with a short and a long link
between them) is eliminated as shown in Fig. \ref{fig1}, 
and the two sites next to
the removed pair are connected by a single link with an effective
resistance $\tilde r$. 
Applying the reduction rules of resistors in series and in parallel, we obtain
\be
\tilde r=r_{a}+r_{c}+\frac{r_b}{r_b+1}.
\label{reff}
\ee 

Let us denote the chemical distance between site $i$ and $i+1$  by
$\ell_i$. Initially, $\ell_i=1$ for all $i$ and when the 
elimination step described above is carried out, 
the chemical distance transforms in a simple way:
\be
\tilde\ell =\ell_{a}+\ell_{c}+1.
\label{leff}
\ee

Now, a renormalization group scheme can be defined, in which
minimal loops are eliminated one after the other as described above, 
exactly in the same order as the long links of loops were created 
in the construction procedure. 
As a consequence, the construction procedure and the renormalization can be
performed simultaneously: once a pair of neighboring active sites is
selected for getting connected it is immediately eliminated and
replaced by an effective short link with resistance and chemical
distance calculated according to Eq. (\ref{reff}) and
Eq. (\ref{leff}), respectively.  
Let us start the construction-renormalization procedure with an infinite
one-dimensional lattice ($N\to\infty$) and assume that the length scale $\xi$, 
which is the inverse of the number
density $c$ of active sites, is large, i.e. $\xi\gg 1$. 
Then the typical resistance and chemical distance
on effective short links 
scale with the length as $r(\xi)\sim\xi^{\zeta}$ and  
$\ell(\xi)\sim\xi^{d_{\rm min}}$, respectively.        
For large $\xi$, the effective resistances and chemical distances are
also large and the transformation rules read asymptotically as 
$\tilde r\simeq r_{a}+r_{c}$ and $\tilde\ell\simeq\ell_{a}+\ell_{c}$,
where the relation $f\simeq g$ is meant as
$\lim_{\xi\to\infty}f/g=1$. 
As the transformation rules of $r$ and $\ell$ are asymptotically
identical, we conclude that for $k=1$
\be 
\zeta =d_{\rm min}.  \qquad \qquad (k=1)            
\ee 
Thus for $k=1$, both $d_{\rm rw}$ and $d_{\rm min}$ are determined by the
resistance exponent $\zeta$. 
This exponent also has a further geometrical meaning for $k=1$.
Let us consider a network built on an open lattice
and call a short link {\it backbone link} 
if its removal results in that the network becomes disconnected.
Taking into account that shortest paths do not contain turnbacks (at
least in the interior of the path) and they 
are composed of long links and backbone links alternately, one can see 
that the fraction of backbone links in a network of size $N$ is
proportional to $N^{d_{\rm min}}$ for large $N$. In other words, the
set of backbone links is a fractal object characterized by the fractal
dimension $d_{\rm min}$.  

\subsection{Uniform model}

Perhaps the simplest model in the class $k=1$ is obtained when 
pairs of neighboring active sites are selected
equiprobably in the course of the construction procedure. 
We call this model the ($k=1$) uniform model. 

The resistance exponent of this network can be calculated as follows.
Consider an infinite system ($N\to\infty$) and assume that, at some
stadium of the renormalization procedure, the number
density of active sites is changed by an infinitesimal amount
$dc<0$.
The differential of the ``resistance density'' 
$\rho\equiv\overline{r}c$  is then 
$d\rho=cd\overline{r}+\overline{r}dc$. On the other hand, 
$d\rho \simeq\overline{r}dc/2$, where the factor $1/2$ comes from that 
in an elimination step two sites are deleted but the total sum of resistances 
is reduced only by $\overline{r}$ on average. 
Combining these equations, we obtain
$cd\overline{r}\simeq-\overline{r}dc/2$, the integration of  
which results in the asymptotical relation
\be 
\overline{r}\sim c^{-1/2}=\xi^{1/2}.
\ee 
Thus, in the $k=1$ uniform model, the dimensions under study are 
$\zeta=d_{\rm min}=1/2$ and $d_{\rm rw}=3/2$. As can be seen, 
random walk is super-diffusive in this network. 

We mention that the calculation carried out above remains valid  
also in the general case when the jump rates are random variables 
provided the expected value of
the inverse jump rates (i.e. resistances) exists.
Thus we obtain the same random-walk dimension for such a disordered 
model, which shows that $d_{\rm rw}$ is determined solely by the structure of the
network. 
 
\subsection{Closest-neighbor networks}

Another possibility to create networks with $k=1$ is when pairs are
not selected with a uniform probability but always the pair (or pairs) 
of actually closest active sites are connected. 
These networks will be termed closest-neighbor networks. 
Here, distances between adjacent sites are rendered initially unequal; 
they are either random or are modulated according to some aperiodic
sequence. Note that a periodic arrangement of short links of different lengths
either would not produce arbitrarily long links or would 
result in equal distances between active sites at some stadium
of the construction procedure, after which the procedure 
would no longer be unambiguous.

\subsubsection{Aperiodic networks}
\label{aper}

\begin{figure}[h]
\includegraphics[width=1\linewidth]{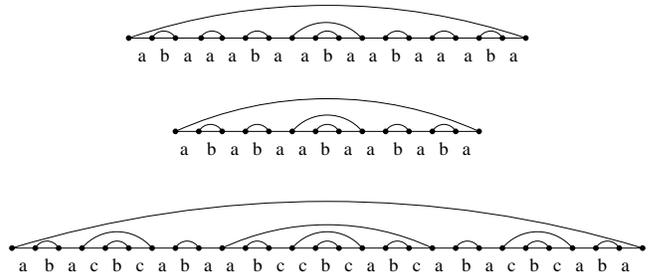}
\caption{\label{fig2} Fragments of the silver-mean, the Fibonacci and
  the tripling network (from top to bottom).}
\end{figure}

First, we consider networks where the short links of different 
lengths are arranged according to aperiodic sequences.
These aperiodic sequences are composed of letters taken from a finite 
alphabet $\{a,b,c,\dots\}$ and are generated by the repeated application of an
inflation rule, which assigns a word (i.e. a finite sequence of
letters) to each letter. 
For instance, the simplest sequence that is suitable for our purposes
is the so-called silver-mean sequence. It is composed of two different
letters, $a$ and $b$, and is generated by the inflation
rule $a\to w_a=aba$, $b\to w_b=a$. Starting from letter $a$, the first
few iterations are $a$,$aba$,$abaaaba$,$abaaabaabaabaaaba$, etc. 
To these finite strings of letters, finite open lattices
can be associated in which the two different edge lengths 
$l_a$ and $l_b(<l_a)$ between adjacent sites 
follow the same sequence as the letters in the strings. 
In networks constructed in this way, there are multitudes of long
links of equal length. These long links will be termed links
of the same generation.    
In general, the aperiodic sequences that we need have to meet 
the following requirements.
First, in order to avoid ambiguities in the construction procedure, 
adjacent short links of shortest length must not emerge.      
Second, the infinite system must be self-similar. 
That means when a new generation 
of long links is formed, the sequence of new distances has  
to follow the original aperiodic sequence.
Third, the order of distances must remain invariant when a
new generation of links is formed, i.e. if $l_a>l_b$ holds, the new
distances must satisfy $\tilde l_a>\tilde l_b$.        
If these requirements are fulfilled, a renormalization step in which
a complete generation of minimal loops is
eliminated corresponds to a reversed inflation step.  
After the sporadic inventions of such sequences in the field of
aperiodic quantum spin chains \cite{hv}, an infinite
class was introduced with the
purpose of studying entanglement entropy in those systems \cite{jz}. 
In the inflation rule of these sequences, the words are composed of 
an odd number of letters and letter $b$, which will represent shortest
links by convention, stands at even places.  
The inflation rule of two-letter sequences with the above properties
can be written in the general form
\be
\sigma_{mn}: \left\{
\begin{array}{c}
{ a \to w_a=aba(ba)^{m-1} }\\
{ b \to w_b=a(ba)^{n-1}, }  
\end{array}
\right. 
\label{2letter} 
\ee
where $n$ and $m$ are integers fulfilling $1\le n\le m$. 
The special case $m=n=1$ corresponds to the silver-mean sequence
mentioned above while, with the choice $m=n=2$, the well-known
Fibonacci sequence is generated. 
An example of three-letter sequences is the tripling sequence, generated
by      
\be
\sigma_{t}: \left\{
\begin{array}{c}
{ a \to w_a=aba} \\
{ b \to w_b=cbc} \\
{ ~c \to w_c=abc.}   
\end{array}
\right. 
\label{tripling} 
\ee
Here, edge lengths are ordered as $l_b<l_c<l_a$. 
The structure of a few aperiodic closest-neighbor networks is illustrated in
Fig. \ref{fig2}. 
 
We shall show that the shortest-path dimension can be
calculated exactly from the substitution matrix $S$ of 
the underlying sequence. 
The elements $S_{\alpha\beta}$ of this matrix are given by
$S_{\alpha\beta}=n_{\alpha}(w_{\beta})$, where  
$n_{\alpha}(w_{\beta})$ is the number of 
occurrences of letter $\alpha$ in the word $w_{\beta}$.
Considering a string of letters $s_i$ and a column vector $v(s_i)$
with the components $[v(s_i)]_{\alpha}=n_{\alpha}(s_i)$, 
it is easy to see that the application of the inflation transformation
results in a longer string $s_{i+1}$ with the vector 
$v(s_{i+1})=Sv(s_i)$. 
Thus, the asymptotic ratio of lengths of successive
strings obtained in the inflation procedure is given by the largest
eigenvalue $\lambda_+$ of $S$. 
Besides, we need the asymptotical scaling of chemical distances
between neighboring active sites 
when a new generation of links forms. 
The chemical distance $\tilde\ell_{\alpha}$ on an effective short link 
represented by letter $\alpha$ after a new generation of long links has formed 
is related to
the previous chemical distances $\ell_{\alpha}$ ($\alpha\neq b$) as 
\be 
\tilde\ell_{\alpha}=
\sum_{\gamma\neq b}n_{\gamma}(w_{\alpha})\ell_{\gamma}+
n_b(w_{\alpha}).
\label{cdt}
\ee
If the density of active sites is small, the chemical distances
are also large and the last term on the r.h.s. of Eq. (\ref{cdt}) 
is negligible.  
We have thus the asymptotical relation 
$\tilde\ell_{\alpha}\simeq\sum_{\gamma\neq b}n_{\gamma}(w_{\alpha})\ell_{\gamma}$. 
This yields that the chemical distances grow 
in a renormalization step asymptotically 
by a factor $\overline{\lambda}_+$,
i.e. $\tilde\ell_{\alpha}/\ell_{\alpha}\simeq\overline{\lambda}_+$,
$\alpha\neq b$,
where $\overline{\lambda}_+$ is the largest eigenvalue of the matrix 
$\overline{S}$ obtained from
$S$ by deleting the row and column related to letter $b$.
As the length scale $\xi$ grows by the factor $\lambda_+$ in a
renormalization step, we obtain finally that the shortest-path
dimension is given by 
$d_{\rm min}=\frac{\ln\overline{\lambda}_+}{\ln\lambda_+}$
and the random-walk dimension is 
\be
d_{\rm rw}=1+\frac{\ln\overline{\lambda}_+}{\ln\lambda_+}. 
\label{aperdw}
\ee
For the family of networks constructed by using two-letter sequences, we have 
$\lambda_+(m,n)=\frac{1}{2}\left[m+n+\sqrt{(m+n)^2+4(m-n+1)}\right]$ and 
$\overline{\lambda}_+(m,n)=m+1$. 
For example, for the silver-mean network we obtain 
$d_{\rm rw}=1+\frac{\ln 2}{\ln(1+\sqrt{2})}$, for the Fibonacci network
$d_{\rm rw}=1+\frac{\ln 3}{3\ln\phi}$ with the golden ratio 
$\phi=\frac{1+\sqrt{5}}{2}$.  
In the tripling network constructed by using the rule in 
Eq. (\ref{tripling}), the set of backbone links is closely related to
the Cantor set and accordingly, the random-walk dimension is 
$d_{\rm rw}=1+\frac{\ln 2}{\ln 3}$. 
Numerical values of $d_{\rm rw}$ are shown in Table \ref{table1} for a
few two-letter networks. 
\begin{table}[h]
\begin{center}
\begin{tabular}{|l|r|}
\hline m=n=1 (silver mean) & 1.7864... \\
\hline m=n=2 (Fibonacci) & 1.7610... \\
\hline m=n=3  & 1.7623... \\
\hline m=n=4 & 1.7683... \\
\hline m=n=5 & 1.7748... \\
\hline
\end{tabular}
\end{center}
\caption{\label{table1} Random-walk dimension of networks constructed on
  the basis of the two-letter inflation rule in Eq. (\ref{2letter}).} 
\end{table}
It follows from Eq. (\ref{aperdw}) that $1<d_{\rm rw}<2$ for 
the aperiodic networks introduced in this section. 
This means that random walk is super-diffusive in these networks just
as in the uniform model. 
Next, we discuss the bounds of $d_{\rm rw}$ in this class of networks. 
As can be seen from the data in Table \ref{table1}, for the two-letter
networks with $n=m$, the random-walk dimension increases monotonously 
with $m$ (i.e. with the length of words) after the minimum at $m=2$ 
and one can see from Eq. (\ref{aperdw})
that it tends to $2$ in the limit $m=n\to\infty$. 
This tendency is intuitively easy to understand: A fragment of
these networks which corresponds to a single word has periodically 
arranged short links, where the links of shortest length $l_b$ are
located at even places. Therefore, the long links generated in the
interior of such a fragment are of 
limited length ($l_b$) and, as a consequence, diffusion in
such fragments is normal. 
Thus the upper bound of $d_{\rm rw}$ in this family of networks is $d_{\rm rw}=2$
and this value, which is characteristic of normal diffusion, can be
approached arbitrarily closely by choosing sufficiently long words in the
inflation rule. 
Concerning the lower bound of $d_{\rm rw}$, presumably there does not exist
representatives of the family of $k=1$ aperiodic networks which
approach the ballistic limit $d_{\rm rw}=1$ arbitrarily closely. 
In fact, the smallest value of $d_{\rm rw}$ that we found is
$d_{\rm rw}=1+\frac{\ln 2}{\ln 3}\approx1.6309$, which is 
realized in the tripling network. 
The geometry of this network is known to be extremal also with respect 
to the von Neumann entropy of aperiodic quantum spin chains \cite{jz}.          In section \ref{sectionk2}, we shall see that larger $k$ may result in
smaller random-walk dimensions.

\subsubsection{Random closest-neighbor network and related networks}       

Another example for closest-neighbor networks is the one
generated with random initial lengths. 
In the case when the initial distribution of lengths is discrete and 
there may be adjacent short links of shortest length present with finite
probability, the construction procedure is extended with the
additional rule that a pair is randomly selected from those with shortest
distance with uniform probability.

The renormalization group scheme of this network with the asymptotical
rule $\tilde\ell\simeq\ell_a+\ell_c$ 
is formally identical to that arising in the context of a 
model of coarsening introduced in
Ref. \cite{bdg}. For that recursion scheme, 
it has been shown that the exponent $z$
that appears in the relation $\ell(\xi)\sim\xi^z$ between the variable
$\ell$ and the length scale $\xi$
is the zero of a transcendental equation and it has been found 
that $z=0.82492412\cdots$. Thus for the random closest-neighbor
network, we have $d_{\rm min}=z$ and $d_{\rm rw}=1+z$. 

Next, we discuss a variant of the random-closest neighbor network, 
the renormalization of which is 
identical to that of certain quantum spin chains
\cite{fisher} and due to this equivalence, the random-walk dimension can
be exactly determined again.    
Let us consider a one-dimensional lattice of size $N$ with 
three variables at each short link: 
the length $l_n$, the chemical distance $\ell_n$, both are initially
equal to $1$, and an independent, identically distributed random 
variable $\beta_n$ that we call $\beta$-distance. 
Now, a cubic network is generated as follows. 
The pair of active sites with the shortest $\beta$-distance, say
$\beta_b$, is chosen and the two sites are connected by a long link. 
The length and the chemical distance on the new effective
short link produced in the equivalent renormalization step are
calculated ordinarily but the $\beta$-distance has an anomalous transformation
rule: 
\be 
\tilde \beta=\beta_{a}+\beta_{c}-\beta_b,  
\ee
where the indices refer to links as given in Fig. \ref{fig1}.
As the structure of the network constructed in this way 
is identical to that of singlet
bonds in the so-called random-singlet phase of antiferromagnetic
quantum spin chains, we call this network {\it random-singlet network}.   
In Ref. \cite{fisher}, it was shown in the limit $N\to\infty$ 
that, when the density $c$ of
active sites goes to zero, the typical value of the variable $\ell$ scales with the length $\xi=1/c$ asymptotically as 
$\ell(\xi)\sim \xi^{\phi/2}$, where $\phi =\frac{1+\sqrt{5}}{2}$ is the
golden ratio. 
The shortest-path dimension of the random-singlet network is thus 
$d_{\rm min}=\frac{1+\sqrt{5}}{4}\approx 0.8090$, and the random-walk
dimension is $d_{\rm rw}=\frac{5+\sqrt{5}}{4}$.  

\begin{figure}[h]
\includegraphics[width=1.\linewidth]{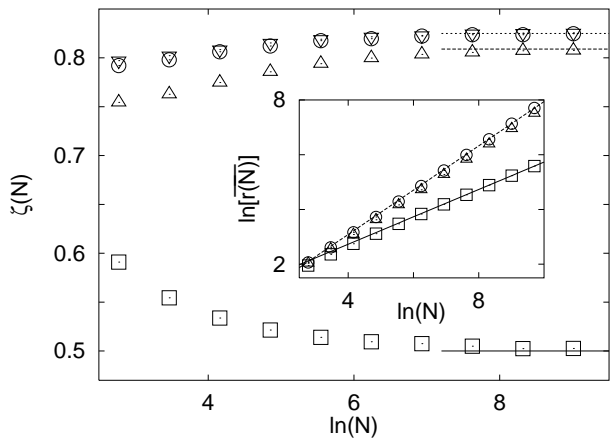}
\caption{\label{k1res}  Effective resistance
  exponent $\zeta(N)\equiv\ln[\overline{r(2N)}/\overline{r(N)}]/\ln 2$
  plotted against the system size for the $k=1$ uniform ({\scriptsize $\Box$}), the random
  closest-neighbor ({\large $\circ$}), the random-singlet ($\vartriangle$) 
  and the random minimal-chemical-distance network ($\triangledown$).
  Horizontal lines indicate the extrapolated values in the limit $N\to\infty$. 
Inset: Dependence of the average resistance on the size $N$.}
\end{figure}

Finally, we examine another variant of the random
closest-neighbor network: 
This is constructed by connecting pairs of active sites 
with the actually shortest chemical distance. 
This means physically that when searching the closest
neighbor of an active site, the already existing long links are also
made use of.  
When generating this network, which we call random
minimal-chemical-distance network, the initial chemical distances are made
random; in the numerical calculations, 
we used the initial values $\ell_n=1+\epsilon_n$, where
$\epsilon_n\ll 1$ is a small random variable. 
One can easily see that the variables $l_n$ and $\ell_n$ 
become positively correlated in the renormalization process of this network, 
i.e. for small values of $l_n$ the variable $\ell_n$ is also typically small.
On the grounds of this observation, we expect that 
the shortest-path dimension of the  
random minimal-chemical-distance network is 
close to that of the random closest-neighbor network. 
According to results of numerical calculations, this is indeed the case. 
We have calculated the effective resistance between endpoints of
networks generated from open lattices of size $N=2^n$, $n=4,\dots,14$. 
Data obtained in this way in $10^6-10^7$ independently generated
networks were averaged for each $N$. Results are shown in
Fig. \ref{k1res}. 
The method was tested on the uniform model, on the random
closest-neighbor network and on the random-singlet network,
for which we obtained $\zeta=0.500(1)$, $\zeta=0.825(1)$ 
and $\zeta=0.809(1)$, respectively. 
The resistance exponent of the random minimal-chemical-distance network
obtained in this way is $\zeta=0.826(1)$, which  
is very close to that of the random closest-neighbor network.

\section{Networks with $k>1$}     
\label{sectionk2}

\begin{figure}[h]
\includegraphics[width=0.5\linewidth]{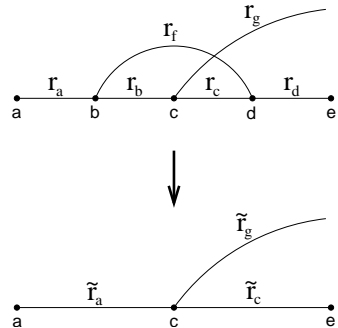}
\caption{\label{rg2} Renormalization scheme for networks with $k=2$.}
\end{figure}

In this section, we will discuss networks with $k>1$, which differ from
the class of networks with $k=1$ studied so far in that 
the resistance exponent and the shortest-path dimension 
are no longer equal. 
We will focus mainly on the case $k=2$, where a renormalization group
scheme similar to that applied for $k=1$ can be formulated. 
Herein, the minimal loops, which are triangles for $k=2$, are
eliminated one after the other as shown in Fig. \ref{rg2}.    
The result of such a renormalization step is that site $b$ and $d$
together with the long link connecting them vanish. 
The new effective resistances can be calculated by employing the
star-triangle transformation of resistor networks. 
This yields 
\beqn
\tilde r_a=r_a+\frac{r_br_f}{r_b+r_c+r_f} \nonumber \\
\tilde r_c=r_d+\frac{r_cr_f}{r_b+r_c+r_f} \nonumber \\
\tilde r_g=r_g+\frac{r_br_c}{r_b+r_c+r_f}. 
\label{DY}
\eeqn
The new feature here compared to the renormalization of $k=1$
networks is that resistances of long links are also transformed.

Note that a similar substitution by which the long links of 
minimal loops are removed without changing the topology of the rest of the 
network cannot be formulated for $k>2$. 

\subsection{The uniform model}

First, the case is considered when pairs of active sites 
are selected randomly with uniform probability in the
construction procedure. 

For $k=2$, when a triangle is eliminated in the renormalization 
procedure,  
the change in the total sum of resistances (including those on long links) 
$\Delta$ can be written in the form 
\be
\Delta=-r_b-r_c+\frac{r_br_c-r_f^2}{r_b+r_c+r_f}.
\label{delta} 
\ee
As can be seen, the average value of $\Delta$ cannot be expressed by the
average values of resistances $r_b$,$r_c$ and $r_f$, therefore 
the evolution of the entire distribution of the latter quantities
should be taken into consideration.   
Another difficulty is that the renormalization rules in
Eq. (\ref{DY}) apparently induce correlations between the quantities 
$\tilde r_a$, $\tilde r_c$ and  $\tilde r_g$.
Therefore, we resorted here to numerical methods again 
in order to estimate $d_{\rm rw}$ and $d_{\rm min}$. 
According to results of numerical renormalization, 
the absolute value of the average of the last term on the r.h.s. of
Eq. (\ref{delta}) is growing proportionally to the average of 
effective resistances in the course of the procedure. 
Although, it is thus not negligible compared to the
average of first two terms on the r.h.s. of Eq. (\ref{delta}), 
it is by an order of magnitude smaller than those two terms. 
If the last term on the r.h.s. of Eq. (\ref{delta}) is omitted,
the resulting simple renormalization process that is 
analytically tractable may provide a first approximation for the
resistance exponent.   
Using that the average resistance of short links is
equal to that of long links, 
which follows after all from the symmetry of the 
transformation rules in Eq. (\ref{DY}), we can write for the 
differential of the resistance density in this simplified process: 
$d\rho =2\overline{r}dc/3$, which leads finally to
$\overline{r}(\xi)\sim \xi^{1/3}$. 

\begin{figure}[h]
\includegraphics[width=0.9\linewidth]{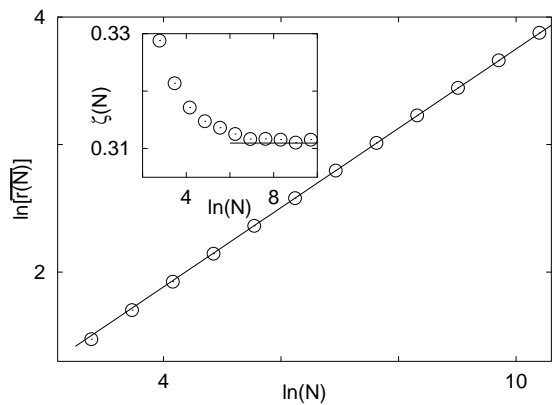}
\caption{\label{k2res} Average resistance of the $k=2$ uniform model
  plotted against the system size. Inset: Effective resistance
  exponent $\zeta(N)\equiv\ln[\overline{r(2N)}/\overline{r(N)}]/\ln 2$
  plotted against the system size. The horizontal line indicates the
  extrapolated value in the limit  $N\to\infty$.}
\end{figure}

We have carried out the renormalization procedure numerically until two
active sites were left in networks generated from periodic chains of
size $N=2^n$, $n=4,\dots,15$. 
The average resistance of the remaining two effective short links 
was calculated from data obtained in  
$10^7-6\times 10^6$ independent networks for each system size $N$.  
The resistance exponent extracted from the finite-size scaling of the
average resistance is $\zeta=0.311(1)$,
see Fig. \ref{k2res}.  

\begin{figure}[h]
\includegraphics[width=1.\linewidth]{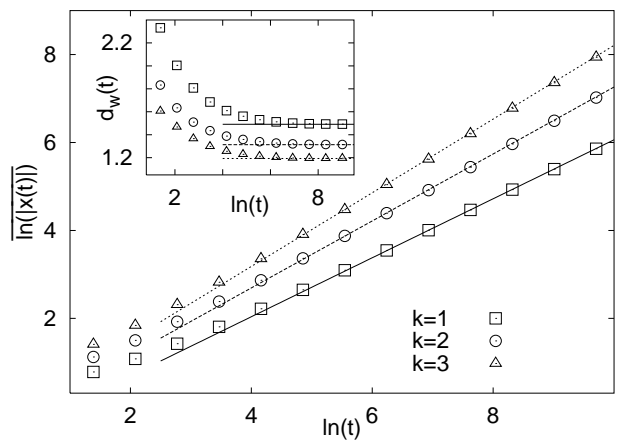}
\caption{\label{sim} Time-dependence of the average logarithmic
  displacement $\overline{\ln |x(t)|}$ of the random walker measured in
  numerical simulations in the uniform model with different values of
  $k$. Inset: Effective random walk dimension $d_{\rm rw}(t)$ calculated from
  neighboring pairs of data points as 
$1/d_{\rm rw}(t)=[\overline{\ln |x(2t)|}-\overline{\ln |x(t)|}]/\ln 2$. 
Horizontal lines indicate the extrapolated values in the limit $N\to\infty$.}
\end{figure}

We have also performed numerical simulations of the random walk in $k=1,2,3$
uniform networks of size $N=10^6$ and measured the displacement $x(t)$ of the
walker at time $t=2^n$, $n=4,\dots,15$. As we measured $x(t)$ in a given
network $N$-times using all the $N$ sites of the network as starting
positions of the walker, the averaging over different stochastic histories 
for a given starting
position was ignored and the quantity $\overline{\ln |x(t)|}$ was
calculated. In addition to the averaging over starting positions, data
obtained in $500$ independent networks were averaged. 
Results are shown in Fig. \ref{sim}. 
The estimated random-walk dimensions
are $d_{\rm rw}(k=1)=1.49(1)$, $d_{\rm rw}(k=2)=1.31(1)$ and $d_{\rm rw}(k=3)=1.19(1)$. 
As can be seen, these values are less accurate compared to the
resistance exponent calculated by numerical renormalization.
Nevertheless, $d_{\rm rw}(k=1)$ measured in this way is compatible with the
exact value $3/2$ and $d_{\rm rw}(k=2)$ is compatible with $\zeta$ 
calculated numerically for $k=2$.      

\begin{figure}[h]
\includegraphics[width=1.\linewidth]{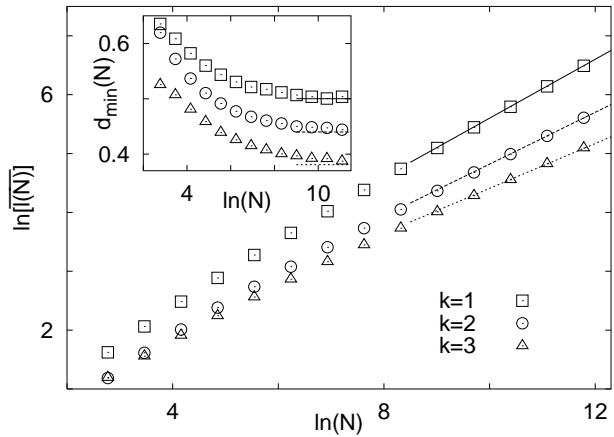}
\caption{\label{diam} Average length of shortest paths
  $\overline{\ell(N)}$ between sites in a distance $l=N/2$  
plotted against $N$ for different values of $k$.   
Inset: Effective shortest-path dimension $d_{\rm min}(N)$ calculated from
  neighboring pairs of data points as 
$d_{\rm min}(N)=\ln[\overline{\ell(2N}/\overline{\ell(N)}]/\ln 2$. 
Horizontal lines indicate the extrapolated values in
  the limit $N\to\infty$.}
\end{figure}

Beside the random-walk dimension, we have measured the shortest-path
dimension for $k=1,2,3$, as well. 
The length of shortest-path between sites in a distance $N/2$ 
was determined by a simple breadth-first search algorithm. 
This was performed in $5\times 10^5-5\times 10^3$ independently
generated networks for each size $N=2^n$, $n=4,\dots,17$, and 
for $N/8$ pairs of sites in each network. 
The average chemical distance $\overline{\ell(N)}$, plotted against
$N$ in Fig. \ref{diam}, was found to grow for large $N$ as 
$\overline{\ell(N)}\sim N^{d_{\rm min}}+C$ with a constant term 
$C\approx -2$, that shifts the effective $d_{\rm min}(N)$
considerably for moderate $N$. 
Using this ansatz, the estimated shortest-path dimensions are:
$d_{\rm min}(k=1)=0.500(1)$, 
$d_{\rm min}(k=2)=0.440(1)$ and $d_{\rm min}(k=3)=0.381(1)$.

\subsection{Regular networks}
\label{aperk2}

In addition to the random networks studied in the previous section, 
one can define regular networks with $k>1$, as well. 
The cubic, hierarchical network studied in Ref. \cite{boettcher} is an
example for a regular network with $k=2$. 
In this section, we shall consider regular $k=2$ networks which
are defined by means of aperiodic sequences. Here, the 
random-walk dimension and the shortest-path dimension can be calculated
exactly. 

Let us consider the subclass of aperiodic sequences discussed in
section \ref{aper}, where the length of words is either one or three. 
One can define networks by using inflation rules with longer words, 
as well, but we shall focus on this simple subclass. 
Once a sequence of this class is given, a finite 
network with odd sites can be defined as follows. 
A finite string of letters which is generated from a single letter is
taken and this time, not the links but the sites of a one-dimensional
lattice are labeled with the letters of the string. 
The sites are grouped into blocks corresponding to words $w_{\alpha}$ in the
inflation rule, which can be done unambiguously. 
Then, sites belonging to one-letter blocks are renamed
according to the reversed inflation rule $w_{\alpha}\to\alpha$, 
where $w_{\alpha}$ is the one-letter word corresponding to the block.      
In blocks composed of three sites, the two lateral sites are
connected, and the middle one is renamed again according to the
reversed inflation rule $w_{\alpha}\to\alpha$, where $w_{\alpha}$ is
the word corresponding to the block.     
The above step is then iterated until only one active site is left.
In finite networks this site is of degree $2$.  

We have not found a general way of calculating $d_{\rm rw}$ and $d_{\rm min}$ in
these networks so they have to be treated individually. 
Two examples will be discussed. 

\subsubsection{The $k=2$ tripling network}

\begin{figure}[h]
\includegraphics[width=1\linewidth]{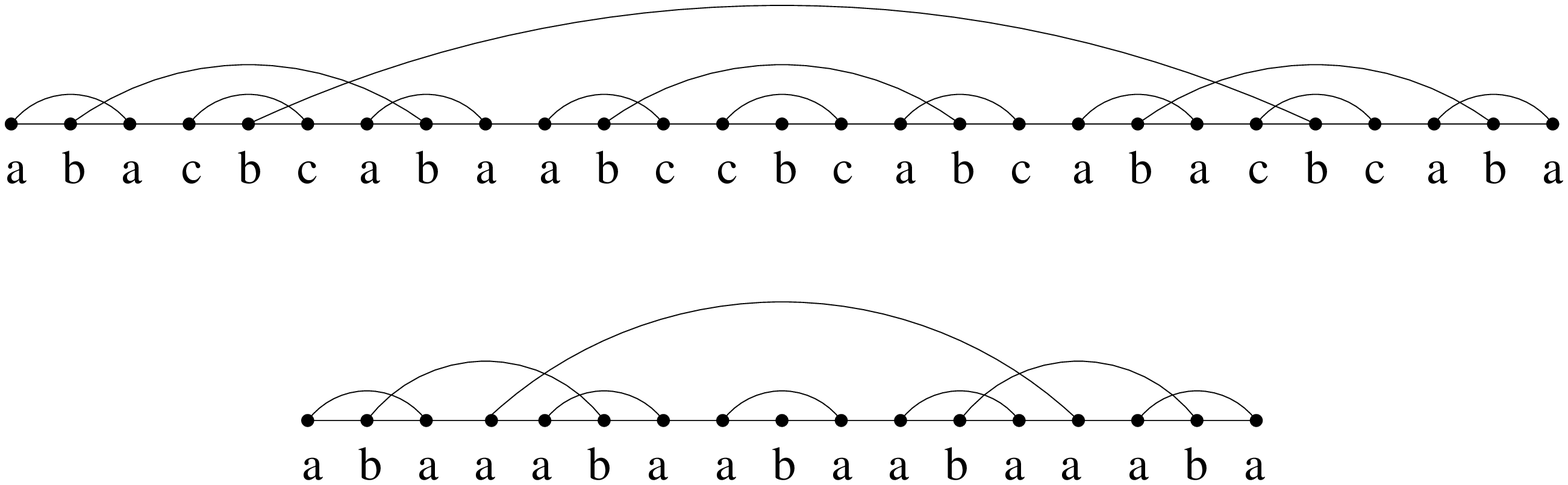}
\caption{\label{k2} A finite $k=2$ tripling network (top)
  and a finite $k=2$ silver-mean network (bottom).}
\end{figure}

First, we examine the $k=2$ tripling network which is constructed by
means of the tripling sequence with the inflation rule in 
Eq. (\ref{tripling}). 
This network is shown in Fig. \ref{k2}. 
Let us generate a sequence of finite strings by applying
the inflation rule $n$-times on letter $a$ and consider the 
corresponding sequence of finite networks built on open lattices 
the sites of which are labeled by these strings.
The size of the $n$th such network is $N_n=3^n$. 
By constructing the first few such networks and observing 
their self-similar structure, 
one can easily check that, for the length of the shortest path between the two
endpoints, the recursive relation $\ell_n=2\ell_{n-1}+1$ holds. 
Thus for large $n$, $\ell_n\sim 2^n$. Expressing $n$ with $N_n$ 
yields $\ell\sim N^{\ln 2/\ln 3}$. The shortest-path
dimension of this network is thus  
$d_{\rm min}=\frac{\ln 2}{\ln 3}\approx 0.6309$.
 
The resistance exponent of this network can be calculated by a 
renormalization procedure in which the generation of long 
links belonging to minimal loops is iteratively eliminated
according to the scheme shown in
Fig. \ref{rg2}. Let us consider the infinite network ($N\to\infty$), 
with resistance $r$ on short links and resistance $p$ on long
links. When the generation of long links belonging to minimal loops is 
eliminated, we obtain a similar network but with modified parameters 
$\tilde r$ and $\tilde p$.   
Using Eq. (\ref{DY}), we obtain for
the renormalized parameters: 
\be 
\tilde r=r+\frac{2rp}{2r+p}, \qquad 
\tilde p=p+\frac{2r^2}{2r+p}.
\ee
These equations yield for the scaling factor of resistances: 
$\overline{\lambda}\equiv\tilde r/r=\tilde p/p=5/3$. 
Keeping in mind that the scaling factor of the length is $\lambda=3$, 
we obtain for the resistance exponent: 
$\zeta=\frac{\ln\overline{\lambda}}{\ln\lambda}=\frac{\ln 5/3}{\ln 3}$
and for the random-walk dimension: 
$d_{\rm rw}=1+\zeta=\frac{\ln 5}{\ln 3}\approx 1.4650$. 

\subsubsection{The $k=2$ silver-mean network}

Next, we discuss the $k=2$ silver-mean network constructed by using
the inflation rule in Eq. (\ref{2letter}) with $m=n=1$ (see Fig. \ref{k2}).
Observing the self-similar structure of finite representatives of this
network, one can formulate the following recursion equation for the
lengths of shortest-paths between endpoints in finite
networks constructed by consecutive strings of letters: 
$\ell_k=3\ell_{k-2}+2$. The asymptotic scaling factor of the
chemical distance is thus $\sqrt{3}$. 
As the asymptotic scaling factor of the size of the
network is $1+\sqrt{2}$ (see Sec. \ref{aper}), the shortest-path dimension is 
$d_{\rm min}=\frac{\ln \sqrt{3}}{\ln(1+\sqrt{2})}\approx 0.6232$.  
 
\begin{figure}[h]
\includegraphics[width=0.8\linewidth]{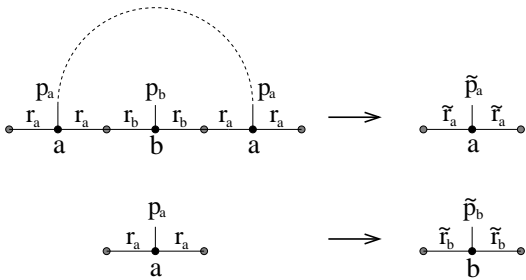}
\caption{\label{rg3} Renormalization scheme for the $k=2$ silver-mean
  network. Sites of the original network are symbolized by black
  circles while the additional sites by grey circles. }
\end{figure}

The calculation of $\zeta$ for the silver-mean network 
is somewhat more involved than for the tripling network.  
In order to find blocks which transform in a self-similar way under a 
renormalization step, an additional site of degree $2$ is inserted in the
middle of each short link. The number of short links (and sites) 
is thus doubled but if we assign a resistance $1/2$ to short
links instead of $1$ in this extended network, 
the effective resistance is obviously equal to that of the original
network. 
Furthermore, instead of keeping track of
resistances of long links, we divide the resistance of a long link
equally among the two sites that it connects. 
This means formally that there is an additional variable $p_i$ at all 
sites of the original network (initially $1/2$), which changes 
in the course of the renormalization procedure, 
just as the resistances of short links. 
After eliminating the first generation of long links, 
the effective resistances of short links
emanating from sites of type $a$ will be different from those 
of short links emanating from sites of type $b$. 
These resistances will be denoted by $r_a$ and $r_b$, respectively. 
The additional variables will also be different at the two types of
sites. These will be denoted by $p_a$ and $p_b$.  
This structure of the parameters remains, however, invariant 
when the subsequent generations of long links are eliminated.   
The substitution of the two different types of building blocks of the
silver-mean network is illustrated in Fig. \ref{rg3}. 
When in an arbitrary stadium of the renormalization procedure the
parameters are $r_a$,$r_b$,$p_a$ and $p_b$ then, after the next
generation of long links is eliminated, the renormalized parameters
can be expressed in terms of the original ones as
\beqn
\tilde r_a=r_a+\frac{p_a(r_a+r_b)}{p_a+r_a+r_b}, \qquad 
\tilde r_b=r_a, \nonumber \\
\tilde p_a=p_b+\frac{(r_a+r_b)^2}{2(p_a+r_a+r_b)}, \qquad 
\tilde p_b=p_a.
\eeqn    
After a lengthy but elementary calculation one obtains that the ansatz 
$\tilde r_a/r_a=\tilde r_b/r_b=\tilde p_a/p_a= \tilde
p_b/p_b\equiv\overline{\lambda}$ solves the above system of equations 
with the scaling factor
$\overline{\lambda}=\frac{1}{1+\sqrt{2}}+\sqrt{2\sqrt{2}}$. 
The resistance exponent is thus 
$\zeta=\frac{\ln\overline{\lambda}}{\ln\lambda}=\ln\left(\frac{1}{1+\sqrt{2}}+\sqrt{2\sqrt{2}}\right)/\ln (1+\sqrt{2})$
and the random-walk dimension of the $k=2$ silver-mean network is 
$d_{\rm rw}=1+\zeta=\ln\left(1+\sqrt{4+2\sqrt{2}}\right)/\ln (1+\sqrt{2})\approx 1.4575$.

\section{Discussion}
\label{disc}  

We have seen that, in the networks studied in this work, 
the distribution of edge lengths is broad and, as a consequence, 
the length of shortest paths grows sub-linearly with the distance
and random walk becomes super-diffusive. 
A model similar to random networks of this class 
(although, not $n$-regular) is 
the model by Benjamini and Berger, where any pairs of sites in a
distance $l$ are connected with probability 
$p_l\sim\beta l^{-s-1}$ for large $l$. 
In that model, the diameter $D$, which is measured in terms of chemical
distances, was conjectured to grow with the system size as 
$D(N)\sim N^{d_{\rm min}(\beta)}$ with some $\beta$-dependent exponent
$0<d_{\rm min}(\beta)<1$ in the marginal case $s=1$ \cite{bb}. 
Later, a power-law lower bound for $\beta<1$ and a 
 power-law upper bound for any $\beta$ were shown to
exist asymptotically with high probability for the size dependence of the 
diameter \cite{coppersmith}.   
Thus, as far as the length of shortest paths (or the diameter) is concerned, 
random networks studied in this paper behave as that model
is conjectured to do. 
Furthermore, we have shown that beside $d_{\rm min}$, 
the random-walk dimension is also influenced by the
details of the structure of networks in the marginal point $s=1$ 
\cite{levy}.   

By means of the class of networks introduced in this work, a
different type of control of the diffusion exponent can be realized 
compared to previous models. 
The control parameter is not the index $s$ as in the L\'evy flight, 
which is set to its marginal value $s=1$ here,
nor the prefactor $\beta$ as in the marginal 
Benjamini-Berger model but the number $k$ and other possible parameters
appearing in the construction procedure. 
The prefactor $\beta$ in the distribution of lengths, as well as 
the shortest-path dimension and the random-walk dimension of 
these networks are non-trivial functions of these parameters. 

We have also shown that by a sequence of $k=1$ aperiodic networks, 
the normal diffusion limit ($d_{\rm rw}=2$) can be approached arbitrarily closely. 
According to numerical results for $k=1,2,3$ uniform networks, the 
random-walk dimension decreases with the parameter $k$ and 
if $k\to\infty$, the latter quantity tends presumably to  
the ballistic limit $d_{\rm rw}=1$. 

As opposed to the diffusion exponent that varies continuously with the index
$s$ in case of the L\'evy flight, the set of possible values of $d_{\rm rw}$ 
which can be realized by these networks is discrete. 
Nevertheless, one can define uniform networks where the parameter 
$k$ is not fixed but more than one values of $k$ are used in the
construction procedure, e.g. $k\in\{n,n+1\}$ with a fixed positive
integer $n$. 
The random-walk dimension of such ``mixed'' networks is expected to
interpolate between that of ``pure'' ones constructed with a
single $k$.   

In a sub-class of networks introduced in this paper, 
the dimensions $d_{\rm min}$ and $d_{\rm rw}$
have been calculated exactly; in some cases they have been estimated
by numerical methods. 
The possible exact calculation of these dimensions for
other networks of this class or finding rigorous bounds on these
quantities is the task of future research.

\appendix
\section{}
\label{app}

Let us consider a network built on an open one-dimensional lattice 
of size $N$.  
Connect the sites at the ends of the lattice (site $1$ and $N$) 
with a directed link which the walker can traverse only in one
direction, say from site $1$ to site $N$ with a unit rate. 
Let us denote the probability that the random walker can be
found at site $i$ in the steady-state by $\pi_i$. 
These probabilities satisfy the set of linear equations 
\be
\sum_{j}w_{ij}(\pi_i-\pi_j)=0,  \quad 1<i<N,
\label{lineq}
\ee
where the summation goes over the set of sites connected with
site $i$ and $w_{ij}$ denotes the jump rate from site $i$ to site $j$; 
in our case $w_{N1}=0$ and $w_{ij}=1$ for other pairs $(i,j)$ of
connected sites. 
Now, one can notice that Eq. \ref{lineq} is analogous to Kirchhoff's
first law for electric circuits. 
Namely, $\pi_i$ plays the role of the potential at
site $i$ and  $1/w_{ij}$ corresponds to the resistance of link $(i,j)$.
Consequently, if the part of the network between site $1$ and $N$ 
(except of the directed link) is replaced
with a single (symmetric) link with a jump rate $1/\tilde r$, where
$\tilde r$ is the effective resistance of the equivalent resistor
network, the steady-state current $J$ through the directed link 
remains unchanged. 
On the other hand, we may write for $J$ in this simplified ``circuit'',
which consists of site $1$ and site $N$, as well as the directed link and the
effective symmetric link connecting them:
\be 
J=\pi_1=(\pi_N-\pi_1)\tilde r^{-1}.
\ee 
If $N\gg 1$, the effective resistance is also large and we have 
$J=\pi_1\approx \pi_N\tilde r^{-1}$.  
If all the links were symmetric, the walker would be
distributed uniformly on the network in the steady state. Although the
link between site $1$ and site $N$ is directed and this leads to that the
network is depleted on the side containing site $1$, the steady-state
probabilities far from the end $1$ such as $\pi_N$ 
are still proportional to $1/N$ for large $N$.     
The expected value of the time $t$ that the walker needs to make a
complete tour on the network (from site $N$ to the same site through
the directed link) is related to the current as $t=1/J$. 
Thus, it scales with the size of the network as 
$t\approx \pi_N^{-1}\tilde r\sim N^{1+\zeta}$ in the large $N$ limit, 
from which we arrive at Eq. (\ref{dwzeta}). 



\begin{thebibliography}{}
\bibitem{levy}
Note that the typical displacement in  
symmetric L\'evy flights with a jump length distribution of the form 
$p_l\sim l^{-2}$ increases with the time as $|x(t)|\sim t$.
\end{thebibliography}
\end{document}